\documentstyle[preprint,aps,prl,multicol,epsfig]{revtex}

\begin{document}

\newcommand{\beq}{\begin{eqnarray}}
\newcommand{\eeq}{\end{eqnarray}}
\renewcommand{\thefootnote}{\fnsymbol{footnote}}

\title{Nonextensive statistical mechanics: Some links with astronomical phenomena \thanks{To appear in the Proceedings of the {\it XIth United Nations / European Space Agency Workshop on Basic Space Sciences}, Office for Outer Space Affairs / United Nations (Cordoba, 9-13 September 2002), ed. H. Haubold, special issue of Astrophysics and Space Science (Kluwer Academic Publishers, Dordrecht, 2003). }}

\author{
Constantino Tsallis  \\
}
\address{
Centro Brasileiro de Pesquisas F\'\i sicas \\
Rua  Xavier Sigaud 150, 
22290-180 Rio de Janeiro, RJ, Brazil
}

\author{
Domingo Prato  \\
}
\address{Facultad de Matematica, Astronomia y Fisica, Universidad Nacional de Cordoba,
 Argentina
}
\author{
Angel R. Plastino \thanks{tsallis@cbpf.br, prato@mail.famaf.unc.edu.ar, vdfsarp9@clust.uib.es}\\
}
\address{Facultad de Ciencias Astronomicas y Geofisicas\\
Universidad Nacional de La Plata and CONICET, 
 CC 727, 1900 La Plata, Argentina\\
and\\
Departament de Fisica, Universitat de les Illes Balears, 07122, Palma de Mallorca, Spain
}
\date{\today}
\maketitle
\begin{abstract}

A variety of astronomical phenomena appear to not satisfy the
ergodic hypothesis in the relevant stationary state, if any. As
such, there is no reason for expecting the applicability of
Boltzmann-Gibbs (BG) statistical mechanics. Some of these
phenomena appear to follow, instead, nonextensive statistical
mechanics. In the same manner that the BG formalism is based on
the entropy $S_{BG}=-k \sum_i p_i \ln p_i$, the nonextensive one
is based on the form $S_q=k(1-\sum_ip_i^q)/(q-1)$ (with
$S_1=S_{BG}$). The stationary states of the former are
characterized by an {\it exponential} dependence on the energy,
whereas those of the latter are characterized by an (asymptotic)
{\it power-law}. A brief review of this theory is given here, as
well as of some of its applications, such as the solar neutrino
problem, polytropic self-gravitating systems, galactic peculiar velocities, cosmic rays and some
cosmological aspects. In addition to these, an analogy with the
Keplerian elliptic orbits {\it versus} the Ptolemaic epicycles is
developed, where we show that optimizing $S_q$ with a few
constraints is equivalent to optimizing $S_{BG}$ with an infinite
number of constraints.

\end{abstract}



\section{introduction}

Connections between dynamics and thermodynamics are far from being
completely clarified. For instance, long-range interactions are
expected to substantially modify various usual thermodynamical
properties. E. Fermi addressed such question with the following words \cite{fermi}: 

{\it The entropy of a system composed of several parts is very
often equal to the sum of the entropies of all the parts. This is
true if the energy of the system is the sum of the energies of all
the parts and if the work performed by the system during a
transformation is equal to the sum of the amounts of work
performed by all the parts. Notice that these conditions are not
quite obvious and that in some cases they may not be fulfilled.
Thus, for example, in the case of a system composed of two
homogeneous substances, it will be possible to express the energy
as the sum of the energies of the two substances only if we can
neglect the surface energy of the two substances where they are in
contact. The surface energy can generally be neglected only if the
two substances are not very finely subdivided; otherwise, it can
play a considerable role.}

In those of L. Tisza \cite{tisza}:

{\it The situation is different for the additivity postulate
$P\;a2$, the validity of which cannot be inferred from general
principles. We have to require that the interaction energy between
thermodynamic systems be negligible. This assumption is closely
related to the homogeneity postulate $P\;d1$. From the molecular
point of view, additivity and homogeneity can be expected to be
reasonable approximations for systems containing many particles,
provided that the intramolecular forces have a short range
character.}

Finally, in those of P.T.  Landsberg \cite{landsberg}:

{\it The presence of long-range forces causes important amendments
to thermodynamics, some of which are not fully investigated as
yet.}

In recent papers, also E.G.D. Cohen \cite{cohen} and M. Baranger
\cite{baranger} have addressed this question.  Indeed, a
significant amount of systems, e.g., turbulent fluids
(\cite{turbulentbeck,turbulentarimitsu} and references therein),
electron-positron annihilation \cite{bediaga}, quark-gluon plasma
\cite{rafelski}, economics
\cite{economics1,economics2,economics3}, motion of {\it Hydra
viridissima} \cite{hydra}, kinetic theory \cite{ademir}, classical
chaos \cite{chaos1}, quantum chaos \cite{chaos2}, quantum
entanglement \cite{entanglement}, anomalous diffusion
\cite{nonlinearFP}, long-range-interacting many-body classical
Hamiltonian systems (\cite{longrange} and references therein),
internet dynamics \cite{internet}, and others, are known nowadays
which in no trivial way accomodate within BG statistical
mechanical concepts. Systems like these have been handled with the
functions and concepts which naturally emerge within nonextensive
statistical mechanics \cite{tsallis1,tsallis2,tsallis3}.

The basic building block of nonextensive statistical mechanics is
the nonextensive entropy \cite{tsallis1} 
\beq 
S_q \equiv
k\frac{1-\sum_{i=1}^{W} p_i^q}{q-1} \qquad (q \in {\cal R}).
\label{sq} 
\eeq 
The entropic index $q$ characterizes the
statistics we are dealing with; $q=1$ recovers the usual BG
expression 
\beq 
S_1=S_{BG}= -k\sum_{i=1}^W p_i \ln p_i \;. 
\eeq 
We
may think of $q$ as a biasing parameter: $q<1$ privileges rare
events, while $q>1$ privileges common events. Indeed, $p<1$ raised
to a power $q<1$ yields a value {\it larger} than $p$, and the
relative increase $p^q/p=p^{q-1}$ is a {\it decreasing} function
of $p$, i.e., values of $p$ closer to 0 (rare events) are
benefited. Correspondingly, for $q>1$, values of $p$ closer to 1
(common events) are privileged. Therefore, the BG theory (i.e.,
$q=1$) is the unbiased statistics. A concrete consequence of this
is that the BG formalism yields {\it exponential} equilibrium
distributions (and time behavior of typical relaxation functions),
whereas nonextensive statistics yields (asymptotic) {\it
power-law} distributions (and relaxation functions). Since the BG
exponential is recovered as a limiting case, we are talking of a
{\it generalization}, not an alternative.

To obtain the probability distribution associated with the
relevant stationary state (thermal equilibrium or metaequilibrium)
of our system we must optimize the entropic form (1) under the
following constraints \cite{tsallis1,tsallis2}: 
\beq 
\sum_ip_i=1
\;, \eeq
 and 
\beq 
\frac{\sum_i p_i^q E_i}{\sum_ip_i^q} =U_q\;,
\eeq 
where $\{E_i\}$ is the set of eigenvalues of the Hamiltonian
(with specific boundary conditions), and $U_q$ is a fixed and {\it
finite} number. This optimization yields 
\beq
p_i=\frac{[1-(1-q)\beta_q (E_i-U_q)]^{1/(1-q)}}{Z_q} \;, 
\eeq
where 
\beq Z_q \equiv \sum_j [1-(1-q)\beta_q (E_j-U_q)]^{1/(1-q)}
\;, \eeq
and \beq \beta_q \equiv \frac{\beta}{\sum_j p_j^q} \;, \eeq
$\beta$ being the optimization Lagrange parameter associated with
the generalized internal energy $U_q$. Equation (5) can be
rewritten as \beq p_i \propto [1-(1-q)\beta^\prime E_i]^{1/(1-q)}
\equiv e_q^{-\beta^\prime E_i} \;, \eeq where $\beta^\prime$ is a
renormalized inverse ``temperature", and the {\it $q$-exponential
function} is defined as $e_q^x \equiv [1+(1-q)
x]^{1/(1-q)}=1/[1-(q-1) x]^{1/(q-1)}$ (with $e_1^x=e^x$).  This
function replaces, in a vast number of relations and phenomena,
the usual BG factor. In particular, the ubiquitous Gaussian
distribution $\propto e^{-ax^2}$ becomes generalized into the
distribution $\propto e_q^{-a_q x^2} = 1/[1+(q-1) a_q
x^2]^{1/(q-1)}$ (fat-tailed if $q>1$, and with a cutoff if $q<1$).

\section{Analogy with the Ptolemy-Kepler problem}

Ancient Greeks believed, through philosophical arguments, that the
orbit of any celestial body ought to be a ``perfect" geometrical
figure, namely a {\it circle}. This belief was very consistent
with the observations of the motion of the stars. Not so, however,
with the observations of the motion of the planets (a word whose
etymology precisely is {\it wanderer}). This nontrivial problem
was attacked by Ptolemy. He introduced, with sensible success at
that time, the idea that the motion of a planet would be that of a
circle rotating around another circle, in turn rotating around
another circle, and so on. The idea of such {\it epicycles} was
recursively used by most astronomers along the centuries over
large regions of the civilized world \cite{henry}. Before Kepler's time, some astronomers had used up to several dozens(!) of
epicycles one on top of the other, in order to increase the
precision of their calculations. Then Kepler arrived and proposed
that the orbit of a planet around the sun would be that of a
geometrical figure which generalizes the circle, namely that of an
{\it ellipse}. This figure needs, for being characterized, one new
parameter, namely the eccentricity $\epsilon$, in addition to say
its largest diameter. The circle is thus recovered as the
particular case $\epsilon =0$.
In very few years, practically all the astronomers of Europe
abandoned the cumbersome calculations with the Ptolemaic
epicycles, and adopted the Keplerian ellipses. Ptolemy idea was
nevertheless not wrong at all! Indeed, the planetary elliptic
motion can be described in terms of infinite series, whose terms
can be identified with the Ptolemaic epicycles. In other words,
the elliptic keplerian motion can be regarded as resulting from an
``infinite series of epicycles". A very lucid discussion of the
correspondence between (i) the series expansion of the elliptic
motion and (ii) the Ptolemaic geometric construction, was provided
by Hoyle \cite{hoyle}. The alluded series for the planetary
elliptic motion can be found in any text-book of Celestial
Mechanics \cite{damby}, and are given (in polar coordinates, as a
function of time $t$) by

\beq \theta = nt + 2\,\epsilon \,sin\, nt +\frac{5}{4}\,
\epsilon^2\, sin \,2nt +... \eeq

\noindent
and

\beq \frac{r}{a} = 1-\epsilon\, cos \,nt +\frac{1}{2}\,
\epsilon^2\, (1- cos \, 2nt) + ... \; \eeq

\noindent
 where $a$ is the semi-major diameter, and $n\equiv
2\pi/P$, $P$ being the so called {\it sidereal period}. If the
orbit is a circle, i.e., $\epsilon =0$, then each polar coordinate
has only one term in its expansion. For any value of $\epsilon \ne
0$, an {\it infinite} number of terms must be used if one wants
the exact Keplerian answer. This sudden change in the number of
terms of course reflects the equally sudden change of symmetry
when one goes from the circle to the ellipse. In other words, if
one wants to stick to circles, that is perfectly possible and
correct, although the price to be paid is that one has to consider
an {\it infinite number of circles}. The other, simpler,
possibility, is of course to use {\it only one ellipse}. One has
however to know, through any accessible procedure, the value of
$\epsilon$ for that particular planet.

Let us go back now to the problem of the use and status of the nonextensive entropy $S_q$ as given in Eq. (1). How universal is the BG entropy? How $S_q$ fundamentally relates to it? What is the status of the entropic index $q$, which characterizes universality classes of nonextensivity? (The most famous one being of course the $q=1$ universality class). Let us address this epistemologically interesting point through the canonical ensemble associated with conservative Hamiltonian systems.

Let us first develop the {\it $q$-logarithm function} 
\beq \ln_qx
\equiv \frac{x^{1-q}-1}{1-q}=\frac{e^{(1-q) \ln x}-1}{1-q}= \ln x
- \frac{q-1}{2} \ln ^2  x + \frac{(q-1)^2}{6} \ln ^3 x + ... \;,
\eeq 
as well as 
\beq p_i^{ q-1} = e^{(q-1)\ln p_i} = 1+ (q-1) \ln
p_i + \frac{(q-1)^2}{2} \ln ^2 p_i + ... \eeq
 The functional to be
optimized for the canonical ensemble is 
\beq \Phi_q(\{p_i\})\equiv
\frac{S_q(\{p_i\})}{k} +\alpha \; \sum_i p_i +\beta \;
\frac{\sum_i p_i^q E_i}{\sum_ip_i^q}=\langle \ln_q (1/p_i) \rangle
+   \alpha \; \sum_i p_i + \beta \; \frac{\langle p_i^{q-1}E_i
\rangle}{\langle p_i^{q-1} \rangle} \;, \eeq 
where $\alpha$ and
$\beta$ are Lagrange parameters, and $\langle ... \rangle \equiv
\sum_ip_i(...)$. We use now the series (11) and (12) inside Eq.
(13), and obtain 
\beq
\Phi_q(\{p_i\})= \frac{S_{BG}(\{p_i\})}{k} - \frac{q-1}{2} \langle \ln ^2 p_i \rangle
- \frac{(q-1)^2}{6} \langle \ln ^3 p_i \rangle +... + \alpha \; \sum_i p_i  \nonumber \\
+ \beta \langle E_i \rangle
+ \beta (q-1)[\langle E_i \ln p_i \rangle -
\langle E_i\rangle \langle \ln p_i\rangle] \nonumber \\
+ \beta \frac{(q-1)^2}{2} 
[\langle E_i \ln ^2 p_i \rangle   -      \langle E_i \rangle\langle \ln ^2 p_i \rangle +2 \langle E_i \rangle \langle \ln p_i \rangle^2 
-2\langle E_i \ln p_i \rangle \langle\ln p_i \rangle]
+...\\
=\Phi_1(\{p_i\}) \nonumber \\
- \frac{q-1}{2} \langle \ln ^2 p_i \rangle -
\frac{(q-1)^2}{6} \langle \ln ^3 p_i \rangle +...
+ \beta (q-1)[\langle E_i \ln p_i \rangle -
\langle E_i\rangle \langle \ln p_i\rangle] \nonumber \\
+ \beta \frac{(q-1)^2}{2} [\langle E_i \ln ^2 p_i \rangle   -      \langle E_i \rangle\langle \ln ^2 p_i \rangle +2 \langle E_i \rangle \langle \ln p_i \rangle^2 
-2\langle E_i \ln p_i \rangle \langle\ln p_i \rangle]
+... 
\eeq 
This optimization is the
usual one within BG statistical mechanics if $q=1$, having a {\it
few constraints}. It suddenly becomes a quite complex one, with an
{\it infinite number of constraints},  if $q\ne1$. Let us further
analyze the generic case.

Let us consider the following generic optimization problem on the
basis of the $S_{BG}$ entropy. The constraints to be used are Eq.
(3), as well as the following ones: 
\beq \langle \ln ^k p \rangle
= C_k \;\; (k=2,3,4,...)\;, 
\eeq where $\{C_k\}$ are fixed finite
quantities, to which we shall associate the Lagrange parameters
$\{\gamma_k\}$. In addition to these constraints, we also have the
following ones: \beq \langle E_i \rangle = D_1\;, 
\eeq \beq
\langle E_i \ln p_i \rangle - \langle E_i\rangle \langle \ln
p_i\rangle = \langle (E_i-\langle E_i\rangle)(\ln p_i-\langle \ln
p_i \rangle)\rangle =D_2\;, 
\eeq 
\beq \langle E_i \ln ^2 p_i \rangle   -      \langle E_i \rangle\langle \ln ^2 p_i \rangle +2 \langle E_i \rangle \langle \ln p_i \rangle^2 
-2\langle E_i \ln p_i \rangle \langle\ln p_i \rangle =D_3 \;, 
\eeq 
etc, where
$D_1,D_2,D_3,..., D_l, ...$ are fixed finite quantities, to which
we shall associate the Lagrange parameters
$\delta_1,\delta_2,\delta_3,..., \delta_l,...$. The optimization
of $S_{BG}$ with all these constraints provides an optimizing
distribution noted $p(E_i; \alpha, \{\gamma_k\}, \{\delta_l\})$.
We shall next consider a particular case of this distribution, the
one where we impose \beq \gamma_k = - \frac{(q-1)^{k-1}}{k!}
\;\;\; (k=2,3,4,...) \;, \eeq \beq \delta_1=\beta \;, \eeq \beq
\delta_2=\beta (q-1) \;, \eeq \beq \delta_3=\beta
\frac{(q-1)^2}{2} \;, \eeq etc. Let us stress that taking a
particular case of  $p(E_i; \alpha, \{\gamma_k\}, \{\delta_l\})$
is perfectly legitimate. [Indeed, as an illustration, let us
remind that in the usual BG statistical mechanics, the
distribution corresponding to the microcanonical ensemble can be
viewed as the $\beta=0$ particular case of the distribution
associated with the canonical ensemble]. Consequently the
stationary distribution becomes a function only of $(E_i,
\beta,q)$. We may alternatively consider that we have defined a
specific new ensemble, whose distribution is fully known once we
know $\beta$ and $q$. The inverse ``temperature" $\beta$ depends
on the physical thermostat with which the system is in contact,
and $q$ depends on the microscopic dynamics of the specific
physical system. The symmetry change that has occurred from $q=1$
to $q \ne 1$ hopefully corresponds to the fact that ergodicity
(homogeneity in phase space) has been lost, and occupancy is now
related to a presumably (multi)fractal or hierarchical geometry in
phase space. The fact that only one new parameter emerges (namely
$q$), and not the infinite number that would correspond to the
entire multifractal function $f(\alpha)$, should be presumably
related to the expectation that the thermodynamical properties
would basically depend only on one number, namely the Hausdorff
dimension of the system (see \cite{lyra} for a connection between
$q$ and the Hausdorff dimension for the simple case of the
logistic-like maps). Although further studies are necessary to
transparently establish the connection of the geometry in phase
space with $q$, we have shown herein that indeed the optimization
of $S_q$ with a few constraints is equivalent to the optimization
of $S_{BG}$ with an infinite number of them. Therefore, it is in
principle possible and correct to use $S_{BG}$ for such complex
systems. The price to be paid is the fact that an infinite number
of constraints must be, in one way or another, taken into account.
The alternative is to use $S_q$ with just a few constraints, but
one needs to know the value for $q$ to be used. The analogy with
the Ptolemy-Kepler problem, where $\epsilon$ must be known, is therefore, in our opinion, quite
striking.

\section{applications}

Let us now briefly review five recent applications of the ideas
associated with nonextensive statistical mechanics to phenomena in astronomy and
astrophysics, namely the solar neutrino deficit \cite{neutrino},
self-gravitating polytropic systems \cite{PP93,P01,polytrope},
peculiar velocities of galaxy clusters \cite{peculiar}, the flux
of cosmic rays \cite{cosmic}, and some cosmological effects
\cite{cosmology}.

\noindent
{\it Solar neutrino problem:}

The solar plasma is believed to produce large amounts of neutrinos
through a variety of mechanisms (e.g., the proton-proton chain).
The calculation done using the  so called Solar Standard Model
(SSM) results in a neutrino flux over the Earth, which is roughly
the {\it double} of what is measured. This is sometimes referred
to as the {\it neutrino problem} or the {\it neutrino enigma}.
There is by no means proof that this neutrino flux defect is due
to a single cause. It has recently been verified that neutrino
oscillations do seem to exist, which would account for part of the
deficit. But it is not at all clear that it would account for the
entire discrepancy. Quarati and collaborators \cite{neutrino}
argue that part of it -- even, perhaps, an appreciable part of it
-- could be due to the fact that BG statistical mechanics is used
within the SSM. The solar plasma involves turbulence, long-range
interactions, possibly long-range memory processes, all of them
phenomena that could easily defy the applicability of the BG
formalism. Then they show \cite{neutrino} in great detail how the
modification of the ``tail" of the energy distribution could
considerably modify the neutrino flux to be expected.
Consequently, small departures from $q=1$ (e.g., $|q-1|$ of the
order of $0.1$ or even less) would be enough to produce as much as $50\%$
difference in the flux. This is due to the fact that most of the
neutrino flux occurs at what is called the {\it Gamow peak}. This
peak occurs at energies quite above the temperature (say $10$ times larger), i.e., at
energies in the tail of the distribution.


\noindent {\it Polytropic Equilibrium Solutions to the
Vlasov-Poisson Equations:}

The first physical application of the non-extensive
 thermostatistical formalism was related to the study of maximum entropy
 solutions to the Vlasov-Poisson equations describing self
 gravitating $N$-body systems like galaxies \cite{PP93,P01}. The
 maximization of the standard Boltzmann-Gibbs entropy under the
 constraints imposed by mass and energy conservation lead to
 the isothermal sphere distribution, which has {\it infinite} mass
 and energy \cite{BT87}. In \cite{PP93,P01}, it was shown that the
 extremalization of the non extensive $q$-entropy under the same constraints
 leads to the stellar polytropic sphere distributions which, for a
 certain range of the $q$ parameter, are endowed with {\it finite}
 mass and energy, as physically expected. This constituted the first clue suggesting that
 the generalized thermostatistical formalism based on $S_q$
 may be of some relevance for the study of systems exhibiting
 non extensive thermodynamical properties due to long range interactions.
 The possible usefulness of non-logarithmic entropic measures in
 the study of stellar systems had also been suggested in \cite{THLB86}.

 Stellar polytropic sphere distributions are of the form

 \begin{eqnarray} \label{polytropic}
 f({\bf x},{\bf v})=
 f({\tilde \epsilon})=& A (\Phi_0 \, - \, {\tilde \epsilon})^{n-3/2}\;\;\;\;&{{\tilde \epsilon}}\leq \Phi_0\\
=&0&{{\tilde \epsilon}}>\Phi_0,\nonumber \end{eqnarray}

\noindent where

\begin{equation} {\tilde \epsilon}=\frac12 {\bf v}^2+\Phi({\bf x}), \label{epsilon}
\end{equation}

 \noindent
 is the total energy (per unit mass) of an individual star,
 and $A$, $\Phi_0$, and $n$ (usually called {\it polytropic index}) are constants. The quantity
 $f({\bf x},{\bf v})d^3x d^3v $ denotes the number of stars with
 position and velocity vectors respectively within the elements
 $d^3x$ and $d^3v$ in position and velocity spaces. The polytropic
 distribution (\ref{polytropic}) exhibits, after an appropriate
 identification of the relevant parameters, the $q$-MaxEnt form.
 The entropic parameter $q$ can be expressed in terms of the index
 $n$ by identifying $n-3/2$ with $1/(1-q)$ (see \cite{P01} for details; notice that here we are identifying $f$ with $p$, and not with the escort distribution $\propto p^q$). We obtain

 \begin{equation} \label{quyene}
 q = \frac{2n-5}{2n-3}.
 \end{equation}
The limit $n \to\infty$ (hence $q=1$) recovers the isothermal sphere case; for $n<5$ 
(hence $q<5/7$), finiteness for mass and energy naturally emerges.  
 Polytropic distributions constitute the simplest, physically
 plausible models for self-gravitating stellar systems \cite{BT87}.
 Alas, these models do not provide an accurate description of the
 observational data associated with real galaxies. In spite of
 this, the connection between the $S_q$ entropy and stellar
 polytropes is of considerable interest. Besides the (in our opinion) notable fact
 that, for a special range of values of $q$, non-extensive thermostatistics leads to
 finite stellar systems, the established connection between the
 $S_q$ entropy and stellar polytropic distributions is interesting for
 other reasons. Polytropic distributions arise in a very natural way
 from the theoretical study of self gravitating systems. The
 investigation of their properties has been of constant interest
 for theoretical astrophysicists during the last one hundred
 years \cite{BT87}. Polytropic distributions are still the focus of an intense
  research activity \cite{polytrope}, and the study of their basic properties
  constitutes a part of the standard syllabus of astrophysics students.
  Now, {\it polytropic distributions happen to exhibit the
  form of $q$-MaxEnt distributions, that is, they constitute
  distribution functions in $({\bf x},{\bf v})$ space that maximize
  the non-extensive functional $S_q$ under the natural constraints imposed
  by the conservation of mass and energy \cite{PP93}} (other
  constraints associated with other conserved quantities may be
  incorporated). Let us recall that the original path leading to
  the $S_q$  entropic form was not motivated by self-gravitating
  systems, nor was it motivated by any
  other {\it specific} system. The entropic form $S_q $ was proposed 
  by recourse to very general arguments dealing with the consideration of
  (i) entropic forms  incorporating power law structures
  (inspired on multifractals) and
  reducing to the standard logarithmic measure in an appropriate
  limit and (ii) the basic properties a functional of the probabilities
  should have in order to represent a physically sensible entropic
  measure \cite{tsallis1}.
  The $q$-entropy is a very natural and, to a certain extent,
  unique generalization of the
  Boltzmann-Gibbs-Shannon measure. Taking this into account, it is
  remarkable that the extremalization of the $q$-measure leads to a
  family of distribution functions of considerable importance in theoretical
  astrophysics. In a sense, we might say that astrophysicists, when studying
  newtonian self-gravitating systems, have been working with $q$-MaxEnt
  distribution functions for a hundred years without being aware of it.
     The discovery of the connection between $S_q$ and
stellar polytropic distributions stimulated the application of nonextensive statistical mechanics 
to the study of other systems with long range
interactions \cite{B96,AT97}. In particular, the analysis by
Boghosian \cite{B96} of metastable states in pure electron plasmas
lead to the identification of the first  maximum $q$-entropy
distribution observed in the laboratory \cite{B96,AT97}. In point
of fact, the mathematical formalism used by Boghosian to describe
the electron plasma is closely related to one associated to the
stellar polytropic distributions \cite{B96}.

\noindent
{\it Peculiar velocities of galaxy clusters}

The COBE (Cosmic Background Explorer) satellite measured the
peculiar velocities (difference of velocity with regard to the
average expansion of the universe)  of some clusters of spiral
galaxies. A distribution was found which exhibits a cutoff around
$500\; Km/s$. The Princeton astrophysical
group \cite{princeton} analyzed the distribution of velocities
within four different cosmological models (cold matter, hot
matter, premieval barionic). None of those attempts succeeded in
reproducing the observed cutoff, although each of those models
involved several free parameters (that were fixed through a
variety of arguments). By assuming within nonextensive statistical
mechanics, an extremely -- almost outrageously -- simplified
model, namely an ideal classical gas!, the empirical velocity
distribution was quite satisfactorily matched \cite{peculiar}.
Only two fitting parameters were used, namely the scale of
velocities and $q \simeq 0.23$. In spite of the extreme simplicity
of the model, the fact that the statistics was allowed to change
proved its high efficiency. The possible primacy of statistics
over models is not new in statistical physics in fact. Even if we
consider the ideal gas hypothesis for a quantum gas, we obtain
acceptably good first approximations for phenomena such as
superconductivity or suprafluidity (using Bose-Einstein
statistics), or such as metallic conductivity of electrons (using
Fermi-Dirac statistics). Of course, more realistic models must include interactions, but an acceptable first approach can indeed be done already  on the basis of an appropriate statistics (even if the model is oversimplified). 

\noindent
{\it Cosmic rays:}

The flux of cosmic rays arriving on Earth is a quantity whose
measured range is among the widest experimentally known ($33$
decades in fact). This distribution refers to a range of energies
which also is impressive ($13$ decades). This distribution is very
far from exponential, as can be verified on Fig 1. This basically indicates that no BG thermal
equilibrium is achieved, but some other (either stationary, or
relatively slow varying) state, characterized in fact by a power
law. If the distribution is analyzed with more detail, one
verifies that two, and not one, power-law regimes are involved,
separated by what is called the ``knee" (slightly below
$10^{16}\;ev$). At very high energies, the power-law seems to be
interrupted by what is called the ``ankle" (close to
$10^{19}\;ev$). One may guess that, along such wide ranges (of
both fluxes and energies), a variety of complex intra- and
inter-galactic phenomena are involved, related to both the sources
of the cosmic rays as well the media they cross before arriving on
Earth. However, from a phenomenological viewpoint, the overall
results amount to something quite simple. Indeed, by solving a
simple differential equation, a quite remarkable agreement is
obtained \cite{cosmic}. This differential equation is the
following one: \beq \frac{dp_i}{dE_i}= -b^\prime p_i^{q^\prime} -b
p_i^q \;. \eeq This differential equation has remarkable
particular cases. The most famous one is $(q^\prime,q)=(1,2)$,
since it precisely corresponds to the differential equation which
enabled Planck, in his October 1900 paper, to (essentially) guess
the black-body radiation distribution, thus opening (together with
his December 1900 paper) the road to quantum mechanics. The more
general case $q^\prime=1$ and arbitrary $q$ is a simple particular
instance of the Bernoulli equation, and, as such, has a simple
explicit solution. This solution has proved its efficiency in a
variety of problems, including in generalizing the Zipf-Mandelbrot
law for quantitative linguistics (see Montemurro's article in
\cite{tsallis3} for a review). Finally, the generic case
$q>q^\prime>1$ also has an explicit solution (though not
particularly simple, but in terms of two hypergeometric functions)
and produces, taking also into account the ultra-relativistic
ideal gas density of states, the above mentioned quite good
agreement with the observed fluxes. Indeed, if we assume
$0<b^\prime<<b$ and $q^\prime<q$, the distribution makes a
crossover from a power-law characterized by $q$ at low energies to
a power-law characterized by $q^\prime$ at high energies, which is
exactly what the cosmic rays exhibit to a quite good
approximation: See Fig. 1.

\noindent
{\it Cosmology:}

Nonextensive statistical mechanics has also been applied to a variety of cosmological and general relativity problems including the cosmic background radiation in a Robertson-Walker universe, the dynamics of inflationary cosmologies, the universal density profile of dark halos, early universe phenomena (e.g., the primordial $^4He$ formation), among others \cite{cosmology}. 

\section{conclusions}

In outer space physics and astrophysics, there is a considerable
amount of anomalous phenomena that require a thermostatistical
treatment which exceeds the usual capabilities of Boltzmann-Gibbs
statistical mechanics. This fact is due to the relevance of
gravitational forces (which are long-ranged), as well as to a
variety of dynamical nonlinear dynamical aspects. Some of these
phenomena appear to be tractable within nonextensive statistical
mechanics, and we have illustrated this with a few typical
examples.

In addition to this, we have argued that, in analogy with the
Ptolemy-Kepler problem concerning the orbits of the planets, the
optimization of the nonextensive entropy $S_q$ with a few
constraints can be equivalently seen as the optimization of the BG
entropy with an infinite number of (ultimately related) constraints. 
The Keplerian orbit is equivalent to {\it infinite} Ptolemaic epicycles, 
and its series expansion in powers of the eccentricity $\epsilon$
abruptly reduces to a {\it single} cercle when  $\epsilon$ precisely 
vanishes. Quite analogously, the 
variational problem associated with $S_q$ for the canonical distribution is 
equivalent to considering $S_{BG}$ with {\it infinite} (related) constraints, and its 
series expansion in powers of $q-1$ abruptly reduces to the $S_{BG}$ variational 
problem with a {\it single} constraint (besides the trivial one associated with 
the norm) when  $q-1$ precisely vanishes

Last but not least, the statements by Fermi and by Tisza that we quoted in the
Introduction constitute a reply which precisely points one of the
various inadvertences contained in Nauenberg's recent criticism of
nonextensive statistical mechanics \cite{nauenbergcritic}. The
nonextensive thermostatistical formalism addresses, among others,
conservative Hamiltonian many-body systems including long-range
interactions. For such systems, the ``surface energy" (using
Fermi's expression), or the ``interaction energy" (using Tisza's expression), is as relevant as the energies of the parts,
and does {\it not} become negligible in the thermodynamic limit,  in contrast with what happens for 
short-range interacting systems. Nauenberg violates precisely this
by imposing his Eq. (7). Moreover, a few lines earlier (his Eq. (5)), he
imposes the factorization of the probabilities. This property is
inconsistent with his own Eq. (7). Indeed, the metaequibrium
distribution of energies is given by the $q$-exponential function
$e_q^x \equiv [1+(1-q)x]^{1/(1-q)}$, and, unless $q=1$,
generically $e_q^{x+y} \ne e_q^x \; e_q^y =e_q^{x+y+(1-q)xy}$.
This type of inconsistency (for finite systems) has already been pointed
out by several authors, starting with \cite{raggio}. It is not
impossible that this specific inconsistency disappears in the
limit $N \to \infty$, if taken before the $t \to\infty$ limit, but
this is highly nontrivial and remains to be proved.

\section*{Acknowledgments}
The possibility of interpreting the optimization of $S_q$ with a
{\it few} constraints as an optimization of $S_{BG}$ with an {\it
infinite} number of constraints emerged during a conversation
between one of us (CT) and M. Gell-Mann, to whom we are very
grateful. We are also indebted to A. Plastino, R.S. Mendes and
E.P. Borges for useful remarks. This work has been partially
supported by PRONEX/MCT, CNPq, and FAPERJ (Brazilian agencies). One of us (ARP) thanks financial support from MECyD (Spanish agency).

\newpage
\begin{figure}[htb]
\begin{center}
\includegraphics[width=0.75\textwidth,keepaspectratio,clip=]{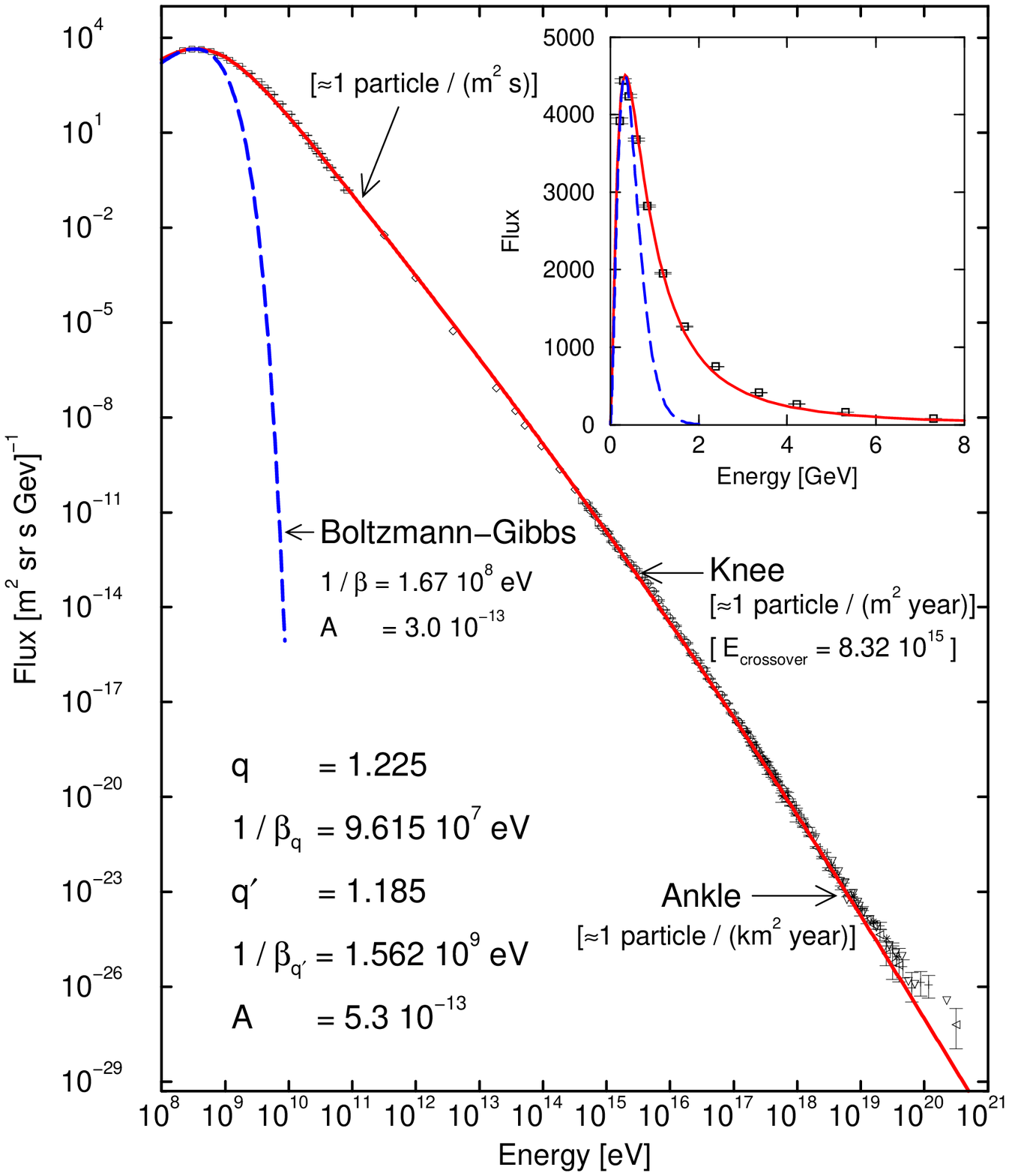}
\end{center}
\caption{\small
Flux of cosmic rays as a function of their energies. See [38] for the origin of the experimental data and theoretical details. The blue dashed curve corresponds to the BG distribution, and the red continuous curve to a crossover between a $q=1.225$ distribution (before the {\it knee}) and a $q^\prime=1.185$ one (after the {\it knee}).}
\label{Fig_1}
\end{figure}



\begin{thebibliography}{10}

\bibitem{fermi}E. Fermi, {\it Thermodynamics} (Dover, New York, 1956).

\bibitem{tisza}L. Tisza, {\it Generalized Thermodynamics}
(MIT Press, Cambridge, Massachussetts, 1966), page 123.

\bibitem{landsberg}P.T. Landsberg,
{\it Thermodynamics and Statistical Mechanics}
 (Oxford university Press, Oxford, 1978), page 102 [also Dover, New york, 1990].

\bibitem{cohen} E.G.D. Cohen, Physica A {\bf 305}, 19 (2002).

\bibitem{baranger} M. Baranger, Physica A {\bf 305}, 27 (2002).

\bibitem{turbulentbeck}C. Beck, G. S. Lewis and H. L. Swinney,
Phys. Rev. E {\bf 63}, 035303 (2001); C. Beck, Phys. Rev. Lett. {\bf 87}, 180601 (2001).

\bibitem{turbulentarimitsu}T. Arimitsu and N. Arimitsu, Physica A {\bf 305}, 218 (2002).

\bibitem{bediaga}I. Bediaga, E. M. F. Curado and J. Miranda, Physica A {\bf 286}, 156 (2000).

\bibitem{rafelski}D.B. Walton and J. Rafelski, Phys. Rev. Lett. {\bf 84}, 31 (2000).

\bibitem{economics1}C. Anteneodo, C. Tsallis and A.S. Martinez,
Europhys. Lett. {\bf 59}, 635 (2002).

\bibitem{economics2}L. Borland, Phys. Rev. Lett. {\bf 89}, 098701 (2002);
Quantitative Finance {\bf 2}, 415 (2002).

\bibitem{economics3}R. Osorio, L. Borland and C. Tsallis,
in {\it Nonextensive Entropy - Interdisciplinary Applications}, M.
Gell-Mann and C. Tsallis, eds. (Oxford University Press, 2003), in
preparation; see also F. Michael and M.D. Johnson, {\it Financial
marked dynamics}, Physica A (2003), in press.

\bibitem{hydra}A. Upadhyaya, J.-P. Rieu, J.A. Glazier and Y. Sawada,
Physica A {\bf 293}, 549 (2001).

\bibitem{ademir}J. A. S. de Lima, R. Silva and A. R. Plastino,
Phys. Rev. Lett. {\bf 86}, 2938 (2001).

\bibitem{chaos1}C. Tsallis, A.R. Plastino and W.-M. Zheng,
Chaos, Solitons \& Fractals {\bf 8}, 885 (1997); U.M.S. Costa,
M.L. Lyra, A.R. Plastino and C. Tsallis, Phys. Rev. E {\bf 56},
245 (1997); M.L. Lyra and C. Tsallis, Phys. Rev. Lett. {\bf 80},
53 (1998); U. Tirnakli, C. Tsallis and M.L. Lyra, Eur. Phys. J. B
{\bf 11}, 309 (1999); V. Latora, M. Baranger, A. Rapisarda, C.
Tsallis, Phys. Lett. A {\bf 273}, 97 (2000); F.A.B.F. de Moura, U.
Tirnakli, M.L. Lyra, Phys. Rev. E {\bf 62}, 6361 (2000); U.
Tirnakli, G. F. J. Ananos, C. Tsallis, Phys. Lett. A {\bf 289}, 51
(2001); H. P. de Oliveira, I. D. Soares and E. V. Tonini, Physica
A {\bf 295}, 348 (2001); F. Baldovin and A. Robledo, Europhys.
Lett. {\bf 60}, 518 (2002); F. Baldovin and A. Robledo, Phys. Rev.
E {\bf 66}, 045104(R) (2002); E.P. Borges, C. Tsallis, G.F.J.
Ananos and P.M.C. Oliveira, Phys. Rev. Lett. {\bf 89},
 254103 (2002); U. Tirnakli, Physica A {\bf 305}, 119 (2002);
 U. Tirnakli, Phys. Rev. E {\bf 66}, 066212 (2002).

\bibitem{chaos2}Y. Weinstein, S. Lloyd and C. Tsallis,
Phys. Rev. Lett. {\bf 89}, 214101 (2002).

\bibitem{entanglement}S. Abe and A.K. Rajagopal, Physica A {\bf 289}, 157 (2001),
C. Tsallis, S. Lloyd and M. Baranger, Phys. Rev. A {\bf 63};
042104 (2001); C. Tsallis, P.W. Lamberti and D. Prato, Physica A
{\bf 295}, 158 (2001); F.C. Alcaraz and C. Tsallis, Phys. Lett. A
{\bf 301}, 105 (2002); C. Tsallis, D. Prato and C. Anteneodo, Eur.
Phys. J. B {\bf 29}, 605  (2002); J. Batle, A.R. Plastino, M.
Casas and A. Plastino, {\it Conditional $q$-entropies and quantum
separability: A numerical exploration}, quant-ph/0207129 (2002).

\bibitem{nonlinearFP}A.R. Plastino and A. Plastino, Physica A {\bf 222}, 347 (1995);
C. Tsallis and D.J. Bukman, Phys. Rev. E {\bf 54}, R2197 (1996);
C. Giordano, A.R. Plastino, M. Casas and A. Plastino, Eur. Phys.
J. B {\bf 22}, 361 (2001); A. Compte and D. Jou, J. Phys. A {\bf
29}, 4321 (1996); A.R. Plastino, M. Casas and A. Plastino, Physica
A {\bf 280}, 289 (2000); M. Bologna, C. Tsallis and P. Grigolini,
Phys. Rev. E {\bf 62}, 2213 (2000); C. Tsallis and E.K. Lenzi, in
{\it Strange Kinetics}, eds. R. Hilfer et al, Chem. Phys.  {\bf
284}, 341 (2002) [Erratum (2002)]; E.K. Lenzi, L.C. Malacarne,
R.S. Mendes and I.T. Pedron, {\it Anomalous diffusion, nonlinear
fractional Fokker-Planck equation and solutions}, Physica A (2003), in press 
[cond-mat/0208332]; E.K. Lenzi, C. Anteneodo and L. Borland, Phys. Rev. E {\bf
63}, 051109 (2001); E.M.F. Curado and F.D. Nobre, {\it Derivation
of nolinear Fokker-Planck equations by means of approximations to
the master equation}, Phys. Rev. E {\bf 67}, 0211XX (2003), in press; C. Anteneodo
and C. Tsallis, {\it Multiplicative noise: A mechanism leading to
nonextensive statistical mechanics}, cond-mat/0205314 (2002).

\bibitem{longrange}C. Anteneodo and C. Tsallis, Phys. Rev. Lett {\bf 80}, 5313 (1998);
V. Latora, A. Rapisarda and C. Tsallis, Phys. Rev. E {\bf 64},
056134 (2001); A. Campa, A. Giansanti and D. Moroni, in {\it Non
Extensive Statistical Mechanics and Physical Applications}, eds.
G. Kaniadakis, M. Lissia and A. Rapisarda,  Physica A {\bf 305},
137 (2002); B.J.C. Cabral and C. Tsallis, Phys. Rev. E {\bf 66},
065101(R) (2002); E.P. Borges and C. Tsallis, in {\it Non
Extensive Statistical Mechanics and Physical Applications}, eds.
G. Kaniadakis, M. Lissia and A. Rapisarda,  Physica A {\bf 305},
148 (2002); A. Campa, A. Giansanti, D. Moroni and C. Tsallis,
Phys. Lett. A {\bf 286}, 251 (2001); M.-C. Firpo and S. Ruffo, J.
Phys. A {\bf 34}, L511 (2001); C. Anteneodo and R.O. Vallejos,
Phys. Rev. E  {\bf 65}, 016210 (2002); R.O. Vallejos and C.
Anteneodo, Phys. Rev. E {\bf 66}, 021110 (2002); M.A. Montemurro,
F. Tamarit and C. Anteneodo, {\it Aging in an infinite-range
Hamiltonian of coupled rotators}, Phys. Rev. E (2003), in press [cond-mat/0205355].

\bibitem{internet}S. Abe and N. Suzuki,  Phys. Rev. E {\bf 67}, 016106 (2003).

\bibitem{tsallis1}C. Tsallis, J. Stat. Phys. {\bf 52}, 479 (1988).

\bibitem{tsallis2}E.M.F. Curado and C. Tsallis, J. Phys. A: Math. Gen. {\bf 24}, L69 (1991)
[Corrigenda: {\bf 24}, 3187 (1991) and {\bf 25}, 1019 (1992)]; C.
Tsallis, R.S. Mendes and A.R. Plastino, Physica A {\bf 261}, 534
(1998).

\bibitem{tsallis3}S.R.A. Salinas and C. Tsallis, eds.,
{\it Nonextensive Statistical Mechanics and Thermodynamics}, Braz.
J. Phys. {\bf 29}, No. 1 (1999); S. Abe and Y. Okamoto, eds., {\it
Nonextensive Statistical Mechanics and its Applications}, Series
{\it Lecture Notes in Physics} (Springer-Verlag, Berlin, 2001); G.
Kaniadakis, M. Lissia and A. Rapisarda, eds., {\it Non Extensive
Statistical Mechanics and Physical Applications}, Physica A {\bf
305}, No 1/2 (Elsevier, Amsterdam, 2002); M. Gell-Mann and C.
Tsallis, eds., {\it Nonextensive Entropy - Interdisciplinary
Applications} (Oxford University Press, 2003), in preparation;
H.L. Swinney and C. Tsallis, eds.,  {\it Anomalous Distributions,
Nonlinear Dynamics, and Nonextensivity}, Physica D (2003), in
preparation. An updated bibliography can be found at the web site
http://tsallis.cat.cbpf.br/biblio.htm

\bibitem{henry} J. Henry {\it Moving Heavens and Earth: Copernicus and the Solar
System} (Icon Books, Cambridge, 2001).

\bibitem{hoyle}F. Hoyle, {\it Nicolaus Copernicus -- An essay on his life and work}
(Harper and Row, Publishers, New York, 1973), page 14.

\bibitem{damby} J.M.A. Danby, {\it Fundamentals of Celestial Mechanics},
Second Edition, (Willmann-Bell Inc., Richmond, 1989).

\bibitem{lyra}F.A.B.F. de Moura, U. Tirnakli and M.L. Lyra, Phys. Rev. E {\bf 62}, 6361 (2000).

\bibitem{neutrino}G. Kaniadakis, A. Lavagno and P. Quarati,
Phys. Lett. B {\bf 369}, 308 (1996); P. Quarati, A. Carbone, G.
Gervino, G. Kaniadakis, A. Lavagno and E. Miraldi, Nucl. Phys. A
{\bf 621}, 345c (1997); G. Kaniadakis, A. Lavagno and P. Quarati,
Astrophysics and space science {\bf 258}, 145 (1998); G.
Kaniadakis, A. Lavagno, M. Lissia and P. Quarati, in Proc. 5th
International Workshop on {\it Relativistic Aspects of Nuclear
Physics} (Rio de Janeiro-Brazil, 1997); eds. T. Kodama, C.E.
Aguiar, S.B. Duarte, Y. Hama, G. Odyniec and H. Strobele (World
Scientific, Singapore, 1998), p. 193; G. Gervino, G. Kaniadakis,
A. Lavagno, M. Lissia and P. Quarati, {\it Non-markovian effects
in the solar neutrino problem}, communicated at "Nuclei in the
Cosmos"  (Volos-Greece, July 1998); M. Coraddu, G. Kaniadakis, A.
Lavagno, M. Lissia, G. Mezzorani and P. Quarti, in {\it
Nonextensive Statistical Mechanics and Thermodynamics}, eds.
S.R.A. Salinas and C. Tsallis, Braz. J. Phys. {\bf 29}, 153
(1999); A. Lavagno and P. Quarati, Nucl. Phys. B, Proc. Suppl.
{\bf 87}, 209 (2000); C.M. Cossu, {\it Neutrini solari e statistica di
Tsallis}, Master Thesis, Universita degli Studi di Cagliari
(2000).

\bibitem{PP93} A. R. Plastino and A. Plastino, {\em Physics Letters
A} {\bf 174}, 384 (1993).

\bibitem{P01}
A.R. Plastino, in {\it Nonextensive Statistical Mechanics and Its Applications}, 
eds. S. Abe and Y. Okamoto (Springer, Berlin, 2001).

\bibitem{polytrope} J.-J. Aly and J. Perez, Phys. Rev. E {\bf 60}, 5185
(1999); A. Taruya and M. Sakagami, Physica A {\bf 307}, 185 (2002);
A. Taruya and M. Sakagami, {\it Gravothermal catastrophe and
Tsallis' generalized entropy of self-gravitating systems II.
Thermodynamic properties of stellar polytrope}, Physica A (2003),
in press [cond-mat/0204315];  P.-H. Chavanis, Astro. and Astrophys. {\bf 386}, 732
(2002); P.H. Chavanis, {\it Gravitational
instability of isothermal and polytropic spheres}, Astro. and
Astrophys. (2003), in press [astro-ph/0207080].

\bibitem{BT87} J. Binney and S. Tremaine, {\it Galactic dynamics}
(Princeton University Press, Princeton, 1987).

\bibitem{THLB86} S. Tremaine, M. H\'enon and D. Lynden-Bell, {\em Mon.
Not. R. Astron. Soc.} {\bf 219}, 285 (1986).

\bibitem{B96} B. M. Boghosian, {\it Phys. Rev. E}, {\bf 53}, 4754 (1996).

\bibitem{AT97} C. Anteneodo and C. Tsallis, {\it J. Mol. Liq.} {\bf
71}, 255 (1997).

\bibitem{princeton}N.A. Bahcall and S.P. Oh, Ap. J. Lett. {\bf 462}, L49 (1996). 

\bibitem{peculiar} A. Lavagno, G. Kaniadakis, M. Rego-Monteiro, P. Quarati
and C. Tsallis, Astrophysical Letters and Communications {\bf  35}, 449 (1998).

\bibitem{cosmic}C. Tsallis, J.C. Anjos and E.P. Borges,
{\it Fluxes of cosmic rays: A delicately balanced anomalous
stationary state}, astro-ph/0203258 (2002).

\bibitem{cosmology}
V.H. Hamity and D.E. Barraco, Phys. Rev. Lett. {\bf  76}, 4664
(1996); V.H. Hamity and D.E. Barraco, Physica A {\bf 282}, 203
(2000); L.P. Chimento, J. Math. Phys. {\bf 38}, 2565
(1997); D.F. Torres, H. Vucetich and A. Plastino, Phys. Rev. Lett. {\bf
79}, 1588 (1997) [Erratum: {\bf 80}, 3889 (1998)]; U. Tirnakli and
D.F. Torres, Physica A  {\bf 268}, 225 (1999); L.P. Chimento, F. Pennini 
and A. Plastino, Physica A {\bf 256}, 197 (1998); L.P.
Chimento, F. Pennini and A. Plastino, Phys. Lett. A {\bf 257}, 275 (1999); D.F. Torres and H.
Vucetich, Physica A {\bf 259}, 397 (1998); D.F. Torres, Physica A {\bf
261}, 512 (1998); H.P. de Oliveira, S.L. Sautu, I.D. Soares and
E.V. Tonini, Phys. Rev. D {\bf 60}, 121301-1 (1999); M.E. Pessah, D.F. Torres
and H. Vucetich, Physica A {\bf 297}, 164 (2001); M.E. Pessah and D.F.
Torres, Physica A {\bf 297}, 201 (2001); C. Hanyu and A. Habe, Astrophys. J. {\bf 554}, 1268 (2001); E.V.
Tonini, {\it Caos e universalidade em modelos cosmologicos com
pontos criticos centro-sela}, Doctor Thesis (Centro Brasileiro de
Pesquisas Fisicas, Rio de Janeiro, March 2002). 

\bibitem{nauenbergcritic}M. Nauenberg,
{\it A critique of non-extensive $q$-entropy for thermal
statistics}, cond-mat/0210561 (2002) [version 2].

\bibitem{raggio}G.R. Guerberoff and G.A. Raggio,
J. Math. Phys. {\bf 37}, 1776 (1996); G.R. Guerberoff, P.A. Pury
and G.A. Raggio,  J. Math. Phys. {\bf 37}, 1790 (1996).

\end{thebibliography}
\end{document}